\documentclass[prd,showpacs,nofootinbib,
]{revtex4}

\renewcommand{\theequation}{\arabic{section}.\arabic{equation}}
\def\lb{\label}
\def\bb{\bibitem}
\def\be{\begin{equation}}
\def\ee{\end{equation}}
\def\ba{\begin{eqnarray}}
\def\ea{\end{eqnarray}}
\def\ds{\displaystyle}
\def\ol{\overline}
\def\rd{{\rm d}}
\def\e{{\rm e}}
\def\nn{\nonumber}
\def\tom{\tilde\omega}
\def\5{_{(5)}}
\def\4{_{(4)}}

\def\2{\sqrt{2}}
\def\3{\sqrt{3}}
\def\6{\sqrt{6}}

\def\A{{\cal A}}
\def\R{{\cal R}}
\begin{document}
\begin{flushright}LAPTH-024/14
\end{flushright}

\title{Bertotti-Robinson and soliton string solutions of $D=5$ minimal
supergravity}
\author{Adel Bouchareb}\email{adel.bouchareb@univ-annaba.dz}
\affiliation{Facult\'{e} des Sciences, D\'{e}ptartement de Physique,
Laboratoire de Physique des Rayonnements, Univ  Annaba, B.P.12,
Annaba 23000, Algeria}

\author{Chiang-Mei Chen} \email{cmchen@phy.ncu.edu.tw}\affiliation{Department
of Physics and Center for Mathematics and Theoretical Physics, \\
National Central University, Chungli 320, Taiwan }
\author{G\'erard Cl\'ement} \email{gerard.clement@lapth.cnrs.fr }
\affiliation{LAPTh, Universit\'e de Savoie, CNRS, 9 chemin de Bellevue, \\
BP 110, F-74941 Annecy-le-Vieux cedex, France}
\author{Dmitri Gal'tsov} \email{galtsov@phys.msu.ru}
\affiliation{Department of Theoretical Physics, Faculty of Physics,
Moscow State University, 119899, Moscow, Russia}

\begin{abstract}
We report on a series of new solutions to five-dimensional minimal
supergravity. Our method applies to space-times with two commuting
Killing symmetries and consists in combining dimensional reduction
on two-spaces of constant curvature with reduction on  a two-torus.
The first  gives rise to various generalized Bertotti-Robinson
solutions supported by electric and magnetic fluxes, which
presumably describe the near-horizon regions of black holes and
black rings (strings). The second provides generating techniques
based on U-duality of the corresponding three-dimensional sigma
model. We identify  duality transformations  relating the above
solutions to asymptotically flat ones  and obtain new globally
regular dyonic solitons.  Some new extremal asymptotically flat
multi-center solutions are constructed too. We also show that
geodesic solutions of three-dimensional sigma models passing through
the same target space point generically split into disjoint classes
which cannot be related by the isotropy subgroup of U-duality.
\end{abstract}


\maketitle

\section{Introduction}
As is well-known, toroidal compactification of multidimensional
supergravities and superstring effective actions to three dimensions
gives rise to gravity-coupled sigma models on symmetric spaces
\cite{ju,BM}. Typically the target space is a coset $G/H$, where $G$
is some semi-simple group combining the manifest geometric
(diffeomorphism and gauge invariance) symmetries of the initial
theory together with its hidden dynamical symmetries, and $H$ is its
isotropy subgroup. This construction joins classical solutions of
the initial theory into duality classes related by the action of $G$
and opens a way to various generating techniques
\cite{kra,Clement:2008qx,revGe}, of which the simplest consists in
generating new solutions acting by $G$ on some seed solution.
Another application is the construction of multicenter solutions
\cite{spat,bps,Bossard:2009at}.

Particularly interesting are solutions associated with geodesic
subspaces of the target space, which arise when the target variables
$\Phi^A$ depend on one (or several) potential functions
$\sigma(x^i)$ realizing a harmonic map between the target space and
the reduced three-space $x^i$ \cite{neukra}. In the one-potential
case, the reduced three-metric is asymptotically Euclidean and the
harmonic function associated with solutions of the black hole type
goes asymptotically to a constant value which can be shifted to
zero: $\sigma(\infty)=0$. Since $\sigma$ plays the role of the
affine parameter on target space geodesics, such solutions can be
seen as geodesics emanating from the point
$X_0=\{\Phi^A[\sigma(\infty)]\}$. Acting on their tangent vectors by
the elements of the isotropy subgroup $H$ leaving the point $X_0$
intact, one can pass from one black hole solution to another with
the same asymptotics. Usually this method is applied to generate
asymptotically flat (AF) solutions by the action of $H$
transformations on a basic seed solution of the Schwarzschild or
Kerr type. For instance, this has been used in $D=5$ to generate
rotating black string solutions to vacuum gravity ($G=SL(3,R)$)
\cite{rasheed} and to minimal supergravity ($G=G_{2(2)}$)
\cite{compere10} from the Kerr black string.

One can also transform geodesic solutions with a given asymptotic
behavior to solutions with different asymptotics, associated with
target space geodesics passing through a different fixed point. For
instance, in $D=5$ black strings and black holes are both
asymptotically flat, but the point at infinity $X_1$ on black hole
geodesics is different from the point at infinity $X_0$ on black
string geodesics. In the case of vacuum gravity one can find
$G$-transformations, not belonging to $H$, which transform $X_0$
into $X_1$, and thus black holes into black strings or vice-versa
\cite{newdil2,giusax}, and this can be extended in principle to the
case of minimal supergravity \cite{Clement:2008qx}. Also of interest
are non-asymptotically flat (NAF) solutions, in particular
Bertotti-Robinson (BR) solutions with AdS asymptotics, which
correspond to near-horizon limits of extremal black holes or black
strings. The transformation between asymptotically flat geodesic
solutions and asymptotically BR solutions has been carried out for
$D=4$ Einstein-Maxwell theory ($G=SU(2,1)$) in \cite{kerr}, and
sketched for $D=5$ minimal supergravity in
\cite{Clement:2008qx,spinem5}. Similar transformations between
geodesic solutions with different asymptotics have also been
discussed in the case of Euclidean Einstein-Maxwell-dilaton-axion
theory ($G=Sp(4,R)$) \cite{emda}. The existence of such $NAF
\leftrightarrow AF$ transformations makes it possible to generate
asymptotically flat solutions from non-asymptotically flat seeds, as
will be demonstrated in the present paper.

Another, quite different question is whether $\it all$ target space
geodesics with the same harmonic potential $\sigma(x^i)$ and passing
through the same fixed point $X_0$ can be related by the action of
the isotropy subgroup $H$. As we shall show on the examples of $D=5$
vacuum gravity and minimal supergravity, the answer is negative. As
a consequence of the existence of a number of invariants which are
preserved by $H$-transformations, the set of geodesics through a
given fixed point, and thus by transitivity the full solution space
(the set of all target space geodesics) splits into disjoint
equivalence classes which cannot be transformed into each other by
$G$-transformations.

Our considerations here will be focussed on five-dimensional minimal
supergravity (MSG5), which attracted special attention in relation
with black rings and black strings  \cite{ER08,compere10}. The
target space of the corresponding three-dimensional sigma model is
the coset $G/H=G_{2(+2)}/((SL(2,R)\times SL(2,R))$ \cite{g2,5to3}.
This theory is a particular case of the more general
Einstein-Maxwell-Chern-Simons theory with arbitrary coupling and
cosmological constant for which we have obtained recently
\cite{reds2} a number of physically interesting NAF solutions via
compactification on two-dimensional constant curvature spaces. This
reduction (with no Kaluza-Klein vectors), recalled in the next
section, leads to a three-dimensional theory possessing non-trivial
solutions of the Banados-Teitelboim-Zanelli (BTZ), self-dual and
G\"{o}del type. Their five-dimensional uplifting gives rise to
generalized Bertotti-Robinson metrics which could serve as
near-horizon limit of yet unknown extremal AdS black rings
\cite{Kunduri:2007vf}. In Sect. 3, we briefly review the sigma model
arising from toroidal reduction of MSG5, and discuss the isotropy
and non-isotropy transformations between geodesic solutions. The map
between the five-dimensional non-asymptotically flat BTZ black
string and the Schwarzchild black string is presented in Sect. 4.

Our main new results stem from the application of the $NAF
\leftrightarrow AF$ map to the non-asymptotically flat G\"odel
string (unrelated to the five-dimensional G\"odel black holes).
First, we show in Sect. 5 that, while the G\"odel string cannot be
$G$-transformed into the Schwarzschild black string, it can be
transformed into the Euclidean Schwarzschild string (the product of
the four-dimensional Euclidean Schwarzschild metric by the time
axis). In Sect. 6, we generate from the G\"odel string a
non-singular (geodesically complete) locally AF metric (with spatial
sections which are asymptotically $R^3\times S^1$), which to our
knowledge is the first exact solution of five-dimensional
supergravity describing a non-BPS regular soliton. This soliton is
supported by electric and magnetic fluxes, endowed with a NUT
parameter and has zero Schwarzchild mass. A further transformation
exchanging the mass and NUT parameters leads to a NUT-less soliton
with positive mass, which is also regular everywhere. Another
unexpected byproduct of our investigation is the existence of a
signature-changing transformation between asymptotically flat
solutions with Lorentzian signature and anti-Euclidean signature
(five timelike coordinates) respectively.

Some of our generalized BR solutions correspond to null geodesics of
the target space, leading to several classes of multicenter
solutions which are presented in Sect. 7. The possibility of
applying our BR-like solutions to the generation of rotating
solutions is discussed in Sect. 8. Our results are summarized in the
closing section.

\setcounter{equation}{0}
\section{Reduction of MSG5 on constant curvature two-spaces}
The bosonic sector of MSG5 is described by the action
 \be\lb{MSG5}
S_5 = \frac1{16\pi G_5}\int \rd^5x \bigg[\sqrt{|g_{(5)}|} \bigg(R\5
- \frac14F\5^{\mu\nu}F_{(5)\mu\nu}  \bigg)    -
\frac1{12\sqrt3}\epsilon^{\mu\nu\rho\sigma\lambda}F_{(5)\mu\nu}
F_{(5)\rho\sigma}A_{(5)\lambda} \bigg]\,,
 \ee
where $F\5 = \rd A\5$, $\mu,\nu,\ldots = 1,\ldots,5$. The sign
convention for the five-dimensional antisymmetric symbol wiil be
fixed throughout this paper by assuming that $\epsilon^{12345} =
+1$, with the space-time coordinates numbered according to their
order of appearance in the relevant five-dimensional metric. The
five-dimensional Maxwell-Chern-Simons and Einstein equations
following from the action (\ref{MSG5}) are
 \ba
\partial_{\mu}(\sqrt{|g\5|}F\5^{\mu\nu}) &=& \frac1{4\sqrt3}
\epsilon^{\nu\rho\sigma\tau\lambda}F_{(5)\rho\sigma}F_{(5)\tau\lambda}\,,
\lb{MCS}\\ {R\5^{\mu}}_{\nu} - \frac12R\5\delta^{\mu}_{\nu} &=&
\frac12F\5^{\mu\rho}
F_{(5)\nu\rho}-\frac18F\5^2\delta^{\mu}_{\nu}\,. \lb{E}
 \ea
Let us assume for the five-dimensional metric and the vector
potential the direct product ansatz\"e
 \be\lb{ans2}
\rd s\5^2 = g_{\alpha\beta}(x^{\gamma})\rd x^{\alpha}\rd x^{\beta} +
a^2 \rd\Sigma_k\,,
 \ee
where $\alpha,\beta,\gamma=1,2,3$, and the two-metrics for $k=\pm 1,
0$ are
 \ba\lb{Sig}
&&\rd\Sigma_1 = \rd\theta^2 + \sin^2\theta \rd\varphi^2,\quad
\rd\Sigma_0 = \rd\theta^2 +  \theta^2 \rd\varphi^2,
\quad\rd\Sigma_{-1}=\rd\theta^2 + \sinh^2\theta \rd\varphi^2\\
&& f_1 = -\cos\theta,\quad f_0= \frac12 \theta^2,\quad
f_{-1}=\cosh\theta
 \ea
with $\varphi\in [0,2\pi]$ and $\theta\in [0,\pi]$ for $k=1$ and
$\theta\in [0,\infty]$ for $k=0,\,-1$.   The vector potential is
decomposed as
 \be
A\5 =A_{\alpha}(x^{\gamma})\rd x^{\alpha} + e f_k\rd\varphi\,.
 \ee
 Here the moduli  $e$ and
$a^2$ are taken to be  constant and real (though for generality we
do not assume outright $a^2$ to be positive).

Following \cite{reds2} one can show that the Eqs. (\ref{E}) reduce
to those following from the  three-dimensional theory
 \be\lb{ac3}
S = \frac{1}{2\kappa}\int \rd^3x \bigg[\sqrt{|g|} \bigg(\R
-\frac14F^{\alpha\beta}F_{\alpha\beta} - 2\lambda\bigg) -
\frac{\mu}4\epsilon^{\alpha\beta\gamma}F_{\alpha\beta}
A_{\gamma}\bigg]
 \ee
with $\kappa=2G_5/|a^2|$ and the identification of parameters:
 \be\lb{lamu}
\lambda=(e^2-4k a^2)/4a^4\,, \quad \mu=g/|a^2|\,,\qquad (g =
2e/\3)\,,
 \ee
provided the three-dimensional scalar curvature is further constrain
by
 \be\lb{constr}
\R = (e^2-6ka^2)/2a^4\,.
 \ee
The three-dimensional theory defined  by the action (\ref{ac3}) is
Maxwell-Chern-Simons electrodynamics coupled to cosmological
Einstein gravity. Several classes of exact stationary solutions to
this theory with constant Ricci scalar are known
\cite{tmgbh,compere06,tmgebh,adtmg,anninos}.

\subsection{BTZ solution} The first class corresponds to neutral
(vacuum) three-dimensional solutions with
 \be\lb{curv}
e^2 = 3ka^2 \,, \qquad \R= 6\lambda = -\frac{3k}{2a^2} \,.
 \ee
The BTZ black hole is a vacuum solution of three-dimensional gravity
with negative $\lambda=-l^{-2}$, so that, for $a^2>0$,
 \be
k=+1\,, \quad a^2 = \frac{l^2}4 \,, \quad e^2 = \frac{3l^2}4\,.
 \ee
Uplifting this to five dimensions according to (\ref{ans2}), we obtain the
following two-parameter family of solutions:
 \ba\lb{bbtz}
\rd s\5^2 &=& -\frac1{2a}(r - Ma)\,\rd t^2 - J\,\rd t\,\rd\,z
+ 2a(r + Ma)\,\rd z^2 \nn\\
&& + a^2\left(\frac{\rd r^2}{r^2 + J^2/4 - M^2a^2}
+ \rd\theta^2 + \sin^2\theta\,\rd\varphi^2\right)\,, \nn\\
A\5 &=& -\sqrt3a\cos\theta\,\rd\varphi
 \ea
($r$ is related to the BTZ radial coordinate $r_{BTZ}$ by $r_{BTZ}^2
= 2a(r+Ma)$). The local isometry group of these solutions is
$SO(2,2)\times SO(3)$.

The solution (\ref{bbtz}) with $z\in R$  coincides with the
decoupling (near-horizon) limit of the general five-dimensional
black string \cite{compere10}. With $z$ periodically identified, the
solution (\ref{bbtz}) may be interpreted as a NAF black ring
rotating along the $S^1$. Moreover, it is the near-horizon limit of
the asymptotically flat black ring with horizon $S^1\times S^2$.

\subsection{Self-dual solutions}
The second class is that of  the ``self-dual'' solutions of
\cite{cam} and \cite{sd} which asymptote to the extreme ($J=Ml$) BTZ
solution (\ref{bbtz}). For these solutions, $F^2=0$ (but
$F_{\alpha\beta}\neq0$), and the constant Ricci scalar has again the
BTZ value $\R=6\lambda\equiv-6l^{-2}$, leading to $l=2a$, $\mu=
\pm2/a$, so that the characteristic exponent $\mu l$ of \cite{cam}
takes the value $\pm4$. The corresponding five-dimensional solution
is:
 \ba\lb{bsd}
\rd s\5^2 &=& \frac1a\left[-(r - aM_{\pm}(r))\,\rd t^2 -
2aM_{\pm}(r)\,\rd t\,\rd z
+ (r + aM_{\pm}(r))\rd z^2 \right] \nn\\
&& \qquad + a^2\left(\frac{\rd r^2}{r^2} + \rd\theta^2 +
\sin^2\theta\,\rd\varphi^2\right)\,, \\
A\5 &=& \3\left[c\left(\frac{r}{a}\right)^{\mp2}(\rd t - \rd z) \mp
a\cos\theta\,\rd\varphi\right]\,, \quad M_{\mu}(r) = M -
\frac{3c^2}{4\pm 1}\left(\frac{r}{a}\right)^{\mp 4} \nn
 \ea
with $c$ a dimensionless parameter.

\subsection{G\"odel solution}
The third class, corresponding to so-called three-dimensional
G\"odel black holes (no relation with the five-dimensional G\"odel
black holes), was given in \cite{compere06} and \cite{tmgebh} (in
the case where the Chern-Simons term for gravity is absent). These
solutions are closely related to the warped $AdS_3$ black hole
solutions of topologically massive gravity
\cite{tmgbh,adtmg,anninos,tmgebh}. Using the notations of
\cite{tmgebh}, the three-dimensional solutions, characterized by a
dimensionless constant $\beta^2 = (1-4\lambda/\mu^2)/2$, have a
constant Ricci scalar $\R = (1-4\beta^2)\mu^2/2$, so that the
constraint (\ref{constr}) implies
 \be\lb{lagod}
\lambda= \frac{5\mu^2}{16}\,,
 \ee
 leading to
 \be\lb{b2l}
\beta^2 = \frac{k}{\mu^2a^2} = k\frac{a^2}{g^2} \,.
 \ee
Comparing (\ref{lagod}) and (\ref{lamu}),  we see that for these
solutions the constant $k$ must be given by
 \be\lb{kga}
k = -\frac{e^2}{6a^2}\,,
 \ee
so that, assuming $a^2>0$, $k=-1$, and $\beta^2=-1/8$.
The resulting five-dimensional solution, derived in \cite{reds2},
may be written in the form
 \ba\lb{gsym}
\rd s\5^2 &=& -(\rd t - gy\,\rd\psi)^2 + \frac{g^2}8\left[
\frac{\rd y^2}{1-y^2} + (1-y^2)\rd\psi^2  + \frac{\rd x^2}{x^2-1} +
(x^2-1)\rd\varphi^2 \right] \nn\\ A\5 &=& -\frac32(\rd t - gy\,\rd\psi)
+\frac{\3}2\,gx\,\rd\varphi\,,
 \ea
with $x^2>1$, $y^2<1$. The local isometry group of this metric is
$SO(2,1)\times SO(2) \times SO(2,1)$. Similarly to the BTZ metric
(\ref{bbtz}), it is geodesically complete.

Let us note that Eq. (\ref{kga}) can also be solved by $k=+1$,
$a^2=-\ol{a^2}<0$ (reduction on a timelike two-sphere). The
resulting ``antiG\"odel'' metric, with the unphysical signature
$(-----)$, may again be written in the form (\ref{gsym}), but with
$g\to-g$, and $x^2<1$, $y^2>1$. Remarkably, as we shall see in Sect.
6, the G\"odel and antiG\"odel solutions, with different spacetime
signatures, can also be transformed into each other by sigma-model
transformations.

\setcounter{equation}{0}
\section{Toroidal reduction and sigma-model transformations}

\subsection{General setup}

All the solutions discussed above admit three commuting Killing
vectors. In this case, beside reduction on a constant curvature
two-surface, one can also carry out toroidal reduction relative to
any two $\partial_a$ ($a=1,2$) of these three Killing vectors,
according to the $GL(2,R)$-covariant Kaluza-Klein ansatz
 \ba\label{st5}
\rd s_{(5)}^2 &=& \lambda_{ab}(\rd x^a + a_i^a\rd x^i)(\rd x^b +
a_j^b\rd x^j) +
\tau^{-1}h_{ij}\,\rd x^i\rd x^j\,, \\
A_{(5)} &=& \sqrt3(\psi_a \rd x^a + A_i\rd x^i)
 \ea
($i,j=3,4,5$) where $\tau = - {\rm det}\lambda$. The Maxwell and
Kaluza-Klein vector fields are then dualized to scalar potentials
$\nu$ (magnetic\footnote{The magnetic potential $\mu$ of
\cite{g2,5to3} is noted here $\nu$ to avoid confusion with the
Chern-Simons coupling constant.}) and $\omega_a$ (twist). In
performing this dualization, we must take care that the scalar
potential $\tau$ can be positive (for most of the solutions
considered here) or negative (in the special case of the
anti-G\"odel solutions with $(5-)$ signature). In this case
$\sqrt{|g\5|} = \varepsilon\tau\sqrt{h}$, where $\varepsilon =$
sign$(\tau)$, and the dualization equations of \cite{g2,5to3} are
modified to
 \be \lb{dualmu} F^{ij} = a^{aj}
\partial^i \psi_a - a^{ai} \partial^j \psi_a + \varepsilon\frac1{\tau \sqrt{h}}
\epsilon^{ijk} \eta_k\,, \qquad \eta_k =
\partial_k \nu +   \epsilon^{ab} \psi_a \partial_k\psi_b
\end{equation}
and
\begin{equation}\lb{dualom}
\lambda_{ab}G^{bij} =  \varepsilon\frac1{\tau \sqrt{h}}
\epsilon^{ijk} V_{ak}\,, \qquad V_{ak} =  \partial_k\omega_a -
\psi_a\left(3\partial_k\nu + \epsilon^{bc}
\psi_b\partial_k\psi_c\right)\,,
\end{equation}
with $G^b_{ij} \equiv \partial_ia^b_j - \partial_ja^b_i$. After
dualization, the reduced field equations derive from the reduced
action (up to a multiplicative constant)
\begin{equation}\lb{sig}
S_3=\int
 \rd^3x\sqrt{h}\left(-R+\frac12G_{AB}\frac{\partial\Phi^A}{\partial x^i}
 \frac{\partial\Phi^B}{\partial x^j}h^{ij} \right),
\end{equation}
where the $\Phi^A$ ($A=1,\cdots,8$) are the eight moduli
$\lambda_{ab}$, $\omega_a$, $\psi_a$, and $\mu$. The action
(\ref{sig}) describes the three-dimensional gravity coupled gauged
sigma model for the eight-dimensional target space with metric:
 \ba\lb{tarmet}
\rd S^2 &\equiv& G_{AB}\rd\Phi^A\rd\Phi^B = \frac12 {\rm
Tr}(\lambda^{-1}\rd\lambda\lambda^{-1}\rd\lambda) +
\frac12\tau^{-2}\rd\tau^2 - \tau^{-1}V^T\lambda^{-1}V \nonumber\\ &&
+ 3\left(\rd\psi^T\lambda^{-1}\rd\psi - \tau^{-1}\eta^2\right) \,,
 \ea
where $\lambda$ is the $2\times2$ matrix of elements $\lambda_{ab}$,
and $\psi$, $V$ the column matrices of elements $\psi_a$,
$V_{a}$.

The target space metric (\ref{tarmet}) admits fourteen Killing
vectors. Nine generate manifest symmetries (generalized gauge
transformations). These belong to several $GL(2,R)$ multiplets: a
four-component mixed tensor ${M_a}^b$ generating $GL(2,R)$ linear
transformations in the $(x^1,x^2)$ plane; a two-component
contravariant vector $R^a$ generating gauge transformations of the
$\psi_a$; another two-component contravariant vector $N^a$ and a
scalar $Q$ generating translations of the dualized potentials
$\omega_a$ and $\nu$. These nine Killing vectors are supplemented by
five Killing vectors $L_a$, $P^a$ and $T$ generating non-trivial
hidden symmetries of the target space. The algebra generated by the
full set of manifest and hidden Killing vectors $J_M$ ($M=1,...,14)$
is that of the fourteen-parameter group $G_{2(+2)}$, and the target
space metric (\ref{tarmet}) is that of the symmetric space
$G_{2(+2)}/((SL(2,R)\times SL(2,R))$. A matrix representative of
this coset can be constructed \cite{g2,5to3} as a symmetric
$7\times7$ matrix $M=M(\Phi)$, given in Appendix A, such that the
target space metric is given  by
 \be \lb{tarmetM}
dS^2 = \frac14{\rm Tr}( M^{-1}dM M^{-1}dM)\,.
 \ee
This form is manifestly invariant under the global action of the
coset isometry group, generating transformation of the moduli
$\Phi\to\Phi'$:
 \be\lb{globalM}
M(\Phi)\to M(\Phi') = P^TM(\Phi)P\,,
 \ee
where the operators $P\in G=G_{2(+2)}$ are generated by the
$7\times7$ matrix representatives $j_M$ of the Killing vectors $J_M$
(also given in Appendix A). These transformations leave invariant
the gravitating sigma model field equations
 \be
\nabla_i\left(M^{-1}\nabla^iM\right) = 0\,,
 \ee
 \be
R_{(3)ij} = \frac14{\rm Tr}( M^{-1}\partial_iM
M^{-1}\partial_jM)\,,
 \ee
where $\nabla_i$ and $R_{(3)ij}$ are the covariant derivative and
Ricci tensor associated with the reduced metric $h_{ij}$. The
$G$-transformations (\ref{globalM}) of the moduli matrix thus belong
to the classical $U$-duality group connecting different solutions
with the same reduced three-metric $h_{ij}$.

\subsection{Geodesic solutions}

All the solutions given in the preceding section admit toroidal
reductions such that the moduli $\Phi^A$ depend
on the three-space coordinates through a single scalar function
$\sigma(x)$. As shown in \cite{neukra}, this potential can be
chosen to be harmonic,
\begin{equation}\label{harm}
\nabla^2\sigma = 0\,,
\end{equation}
so that the field equations reduce to
 \be
\frac{\rd}{\rd\sigma}\left(M^{-1}\frac{\rd M}{\rd\sigma}\right) = 0
\,,
 \ee
 \be\lb{Rijs}
R_{(3)ij}  =  \frac14{\rm Tr}\left(M^{-1}\frac{\rd
M}{\rd\sigma}\right)^2
\partial_i\sigma\partial_j\sigma \,.
 \ee
The first of these equations is the geodesic equation for the
target space metric (\ref{tarmet}) with $\sigma$ the affine
parameter. It is solved by
\begin{equation}\label{geo}
M = \eta{\rm e}^{{\cal A}\sigma}\,,
\end{equation}
where $\eta \in G/H$ and ${\cal A} \in {\rm Lie}(G)-{\rm Lie}(H)$ are
constant matrices, which transform under the action (\ref{globalM})
of $G$ according to
 \be\lb{etaA}
\eta'=P^T \eta \,P\,,\qquad {\cal A}'=P^{-1}{\cal A}\; P\,.
 \ee
The second equation (\ref{Rijs}) then reduces to
\begin{equation}\lb{Rtrac}
R_{(3)ij}=\frac{1}{4}{\rm Tr}({\cal
A}^2)\partial_i\sigma\partial_j\sigma\,.
\end{equation}
The sign of the spatial curvature, hence the nature of the
three-geometry, depends on the sign of the constant ${\rm Tr}({\cal
A}^2)$. This trace is invariant under general $G-$transformations.
If the target space metric has indefinite signature, which for
Lorentzian solutions is the case in presence of vector charges, then
geodesics are split into three disjoint classes: a timelike class
(Tr($\A^2)>0$), which includes black hole solutions; a null class
(Tr($\A^2)=0$), which corresponds to extremal black holes and
multi-black hole solutions; and a spacelike class (Tr($\A^2)<0$),
which includes wormhole solutions. In the present paper, we will be
mainly concerned with solutions of the timelike class (solutions of
the null class will be discussed in Sect. 7). In that case the
constant ${\rm Tr}({\cal A}^2)$ can be fixed to
 \be
{\rm Tr}(\A^2)=4\,.
 \ee
The Einstein-scalar equations (\ref{Rtrac}) then determine the reduced metric
$h_{ij}(x)$ (up to coordinate transformations):
 \be\lb{3red}
\rd s_{(3)}^2 \equiv h_{ij}\,\rd x^i\rd x^j = \rd r^2 +
(r^2-m^2)(\rd\theta^2 + \sin^2\theta\,\rd\varphi^2)\,,
 \ee
and the scalar potential $\sigma(x)$ (up to linear transformations).
 \be\lb{sigf}
\sigma = \ln f\,, \qquad f(x) = \frac{r-m}{r+m}\,.
 \ee

\subsection{Isotropy transformations}
 We can regard geodesic solutions as curves in target space passing
through the point $X_0=\{\Phi^A (x_0)\}$ where $x_0$ is some
characteristic point in the reduced three-space. Often (but not
necessarily), one takes $x_0=\infty$, and chooses for the harmonic
potential $\sigma(x)$ a gauge such that $\sigma(\infty)=0$. In that
case, the constant matrix $\eta = M(\infty)$ specifies the
asymptotic nature of the solution under consideration. In this
paper, we consider only (not necessarily black) string solutions. In
the locally asymptotic Minkowskian (LAM) case (with possible Misner
string singularities), in a gauge where $\lambda(\infty)=$
diag$(-1,1)$ and the other moduli vanish at infinity, the
corresponding matrix $\eta$ is given by
 \be
\eta_S = \mbox{\rm diag} (-1,\; 1,\; -1,\; -1,\; 1,\; -1,\; 1)\,.
 \ee
This is invariant under the transformations $P\in H =
SL(2,R)\times SL(2,R)$ generated by
the eight elements of the isotropy subalgebra
  \be\lb{isos}
h_S = \left\{n^0+\ell_0,\,n^1-\ell_1,\,{m_1}^0+{m_0}^1,\,r^0+p_0,\,
r^1-p_1,\,q-t\right\}\,.
 \ee

Starting from a given geodesic solution, e.g. the Schwarzschild
black string
 \ba\lb{blackS}
\rd s\5^2 &=& - f(r)\,\rd t^2 + \rd z^2 + f^{-1}(r)\left[\rd r^2  +
(r^2-m^2)\rd\Omega_2^2\right]\,,  \nn\\
A\5 &=& 0\,,
 \ea
one can generate other LAM solutions through the action of isotropy
transformations $P \in H_S$. The question arises, whether one can
obtain {\em all} geodesics passing through $X(\infty)$ in this way?

Consider for instance the simple example of the coset
$SL(n,R)/SO(n-2,2)$ (($n+2$)-dimensional vacuum gravity). The
dimensions of the invariance group $G$ and of the isotropy subgroup
$H$ are $n_G = n^2-1$ and $n_H=n(n-1)/2$. The number of charges in
the charge matrix $\A$ (with ${\rm Tr}(\A) = 0$) is equal to the
dimension of the coset $n_c=n_G-n_H=(n+2)(n-1)/2$. However, in a
given equivalence class (under isotropy transformations), the number
of independent charges is lower. The reason is that such
transformations preserve the  $(n-1)$ invariants ${\rm Tr}(\A^2)$,
... , ${\rm Tr}(\A^n)$, so that the number of independent charges is
only $n_c-(n-1)= n(n-1)/2 = n_H$. Furthermore, not all geodesic
solutions with given values $c_i$ ($i = 1, ..., n-1$) of the $(n-1)$
trace invariants are equivalent to some given solution with the same
values for these invariants. The corresponding charges belong to an
$n(n-1)/2$ dimensional variety $V$ which is the intersection of the
$i$-dimensional varieties $Tr(A^{i+1}) = c_{i+1}$ ($i=2,\cdots,n$)
and may have several connected components.

Take the case of five-dimensional Lorentzian vacuum gravity (E5)
reduced to three Euclidean dimensions. The target space is
$SL(3,R)/SO(2,1)$. Consider locally asymptotically flat geodesic
solutions $M = \eta_S \exp[\A\sigma]$, with $\eta_S = {\rm
diag}(-1,1,-1)$. A necessary condition for regularity of these
solutions is ${\rm det}\,\A = 0$. After using the tracelessness and
normalization conditions ${\rm Tr}(\A) = 0$, ${\rm Tr}(\A^2) = 2$,
leading to $\A^3=\A$, the charge matrix $\A$ can be diagonalized to
one of the possible three forms
 \be
\A_1 = {\rm diag}(1, 0, -1)\,, \quad \A_2 = {\rm diag}(0, 1, -1)\,,
\quad \A_3 = {\rm diag}(1, -1, 0)\,,
 \ee
which are inequivalent (cannot be transformed into each other by
similarity transformations belonging to $H = SO(2,1)$). The first
one leads to the class of the Schwarzschild black string
(\ref{blackS}) (S), which is the direct product of the
four-dimensional Lorentzian Schwarzchild black hole by spacelike
$S^1$, and other black strings, as well as black holes. The second
one leads to the class of soliton strings generated from the
Euclidean Schwarzschild string (ES), the direct product of the
four-dimensional Euclidean Schwarzchild solution by the timelike
real axis:
 \be\lb{euclS}
\rd s\5^2 =  - \rd t^2 + f( r)\,\rd z^2  + f^{-1}( r)[\rd r^2 +
(r^2-m^2)\rd\Omega_2^2]
 \ee
(which is also regular if the coordinate $z$ is periodically
identified with suitable period). And the third leads to the class
generated from the singular solution
 \be\lb{sol3}
\rd s\5^2 = - f(r)\,\rd t^2 + f^{-1}(r)\,\rd z^2 + \rd r^2 +
(r^2-m^2) \rd\Omega_2^2\,.
 \ee
So in this case geodesic solutions on a given three-dimensional
reduced metric fall into three distinct equivalence
classes\footnote{The obstruction discussed here is clearly different
from that considered in \cite{horn}, which arises when the charge
matrix $\A$ has complex eigenvalues.}. We will show in the following
that this result holds also for minimal five-dimensional
supergravity.

\subsection{Non-isotropy transformations}

Another question is whether one can transform a solution
corresponding to a geodesic passing through a given point $X$ (for
instance LAM) to a solution corresponding to different asymptotics,
i.e. passing through a different point $X'$ of the target space.
Such transformations $P_{XX'}\notin H$ will lead from $\eta_X$ to
 \be\lb{PXXprime}
\eta_{X'}= P_{XX'}^T\eta_XP_{XX'} \neq \eta_X\,.
  \ee
Three quite different type of transformations are actually
concerned. The transformation $P_{XX'}$ can be a generalized gauge
transformation, which does not modify the intrinsic solution. Or it
can relate asymptotically flat (AF) solutions with intrinsically
different asymptotics, for instance transform black strings into
black holes \cite{newdil2,giusax}. Or finally it can relate AF and
non-asymptotically flat (NAF) solutions, transforming for instance
an AF black hole to another exact solution which is its near-horizon
limit and back \cite{kerr,emda}. In all cases, since the $U$-duality
group acts transitively on the target space, inequivalent geodesics
passing through $X$ will be transformed into inequivalent geodesics
passing through $X'$. Thus, the existence of several distinct
equivalence classes of geodesics through a given point $X$ of target
space actually means that the solution space has several disjoint
components, which cannot be related by invariance group
transformations, irrespective of the asymptotics involved.

\setcounter{equation}{0}
\section{From BTZ to Schwarzschild}
We first give a non-trivial example of relating solutions with
different asymptotics, namely the BTZ ring (\ref{bbtz}) and
Schwarzschild black string (\ref{blackS}). After toroidal reduction
relative to $\partial_t$ and $\partial_z$, the three-dimensional
reduced metric $h_{ij}$ is in both cases (\ref{3red}), with $m^2
\equiv M^2a^2 - J^2/4$ in the BTZ case. This means that the
corresponding $7\times7$ matrix representatives $M_S$
(Schwarzschild) and $M_B$ (BTZ) may be related by a $G_{2(2)}$
transformation,
 \be
M_B = P_{SB}^T M_S P_{SB}\,.
 \ee
To construct the transformation matrix $P_{SB}$, we can use the fact
that both $M_S$ and $M_B$ are geodesic solutions $M =
\eta\,\e^{\A\sigma}$ with $\sigma = \ln f$, so that their asymptotic
and charge matrices $\eta$ and $\A$ are related by
 \ba
\eta_B &=& P_{SB}^T \eta_S P_{SB}\,, \lb{ebs} \\
\A_B &=& P_{SB}^{-1}\A_SP_{SB}\,, \lb{abs}
 \ea
i.e. $P_{SB}$ is the inverse of a similarity transformation $P_{BS}$
bringing the matrix $\A_B$ into the diagonal form $\A_S$, normalized
by the constraint (\ref{ebs}), and subject to additional constraints
ensuring that it belongs to $G_2$. Necessary conditions for the
transformation matrix $P_{SB}$ to belong to $G_2$ are $P_{SB} \in
SO(4,3)$, which implies
 \be\lb{kpk}
P_{SB}^{-1} = KP_{SB}^TK\,, \quad {\rm det}(P_{SB}) = +1\,,
 \ee
where $K$ is the matrix
 \be\lb{K}
K = \left(\begin{array}{ccccccc} 0 & 0 & 0 & 1 & 0 & 0 & 0 \\
0 & 0 & 0 & 0 & 1 & 0 & 0 \\ 0 & 0 & 0 & 0 & 0 & 1 & 0 \\
1 & 0 & 0 & 0 & 0 & 0 & 0 \\ 0 & 1 & 0 & 0 & 0 & 0 & 0 \\
0 & 0 & 1 & 0 & 0 & 0 & 0 \\ 0 & 0 & 0 & 0 & 0 & 0 & -1
\end{array}\right)\,,
 \ee
but these conditions are not sufficient.

Without loss of generality, we consider only in the following the
static case ($J=0$, $M=m/a$). The general solution with $J\neq0$ can
be recovered from this by a Lorentz boost in the 2-Killing space (a
trivial $G_2$ transformation). The reduction of (\ref{bbtz}) leads
to the scalar potentials
 \ba
\lambda &=& \frac1a\left(\begin{array}{cc} -r+m & 0 \\ 0 & r+m
\end{array}\right) \,, \quad \tau = \frac{r^2-m^2}{a^2}\,, \quad
\psi=0\,, \nn\\
\omega&=&0\,,\quad \nu = \frac{r}a\,.
 \ea
From these potentials one constructs, according to the prescriptions
of \cite{g2,5to3}, the $7\times7$ matrix representative
 \ba
&& M_B =
\frac{m^2}{r^2-m^2}\times\nn\\
&&\times\left(\begin{array}{ccccccc} \ds\frac{r-m}a & 0 & 0 & 0 &
-\ds\frac{r(r-m)}{m^2} & 0 & 0 \\ 0 & -\ds\frac{r+m}a & 0 &
-\ds\frac{r(r+m)}{m^2} & 0 & 0 & 0 \\ 0 & 0 & -\ds\frac{a^2}{m^{2}}
& 0 & 0 & -\ds\frac{r^2}{m^{2}} & \sqrt2\ds\frac{ar}{m^{2}}
\\ 0 & -\ds\frac{r(r+m)}{m^2} & 0 & -\ds\frac{a(r+m)}{m^2} & 0 &
0 & 0 \\ -\ds\frac{r(r-m)}{m^2} & 0 & 0 & 0 & \ds\frac{a(r-m)}{m^2}
& 0 & 0 \\ 0 & 0 & -\ds\frac{r^2}{m^{2}} & 0 & 0 &
-\ds\frac{m^2}{a^{2}} & \sqrt2\ds\frac{r}a \\ 0 & 0 &
\sqrt2\ds\frac{ar}{m^{2}} & 0 & 0 & \sqrt2\ds\frac{r}a &
-\ds\frac{r^2+m^2}{m^2}
\end{array}\right)\,.
 \ea
The resulting constant matrices $\eta$ and $\A$ are
 \be\lb{etab}
\eta_B = \left(\begin{array}{ccccccc} 0 & 0 & 0 & 0 & -1 & 0 & 0
\\ 0 & 0 & 0 & -1 & 0 & 0 & 0 \\ 0 & 0 & 0 & 0 & 0 & -1 & 0
\\ 0 & -1 & 0 & 0 & 0 & 0 & 0 \\ -1 & 0 & 0 & 0 & 0 & 0 & 0 \\
0 & 0 & -1 & 0 & 0 & 0 & 0 \\ 0 & 0 & 0 & 0 & 0 & 0 & -1
\end{array}\right)\,,
 \ee
 \be\lb{ab}
\A_B = \frac12\left(\begin{array}{ccccccc} 1 & 0 & 0 & 0 & M^{-1} &
0 & 0 \\ 0 & -1 & 0 & - M^{-1} & 0 & 0 & 0
\\ 0 & 0 & 0 & 0 & 0 & 0 & \2M
\\ 0 & - M & 0 & -1 &
0 & 0 & 0 \\  M & 0 & 0 & 0 & 1 & 0 & 0 \\
0 & 0 & 0 & 0 & 0 & 0 &
\2M^{-1} \\
0 & 0 & \2M^{-1} & 0 & 0 & \2M & 0
\end{array}\right)\,.
 \ee

On the other hand, the matrix representative for the Schwarzschild
black string
 \be\lb{ms}
M_S = \mbox{\rm diag} (-f,\; 1,\; -f^{-1},\; -f^{-1},\; 1,\; -f,\;
1)
 \ee
corresponds to the constant matrices
 \ba
\eta_S &=& \mbox{\rm diag} (-1,\; 1,\; -1,\; -1,\; 1,\; -1,\; 1)\,, \\
\A_S &=& \mbox{\rm diag} (1,0,-1,-1,0,1,0)\,.
 \ea

The procedure outlined above for determining the transformation
matrix $P_{SB}$ does not ensure that it belongs to the group $G_2$.
However, educated guesses show that that this matrix can be written
as the product of two elementary $G_2$ transformations, i.e.
exponentials of $g_2$ generators, as given in matrix form in
\cite{g2} (Appendix A) and \cite{5to3}. First, the transformation
 \be
P_{BS0} = \exp[\alpha_0(q+t)]\,, \quad \alpha_0 = -\pi/4\,,
 \ee
acting bilinearly on $M_B$ transforms $\eta_B$ to $\eta_S$ and
$\A_B$ to
 \be
\A'_B = \frac1{4M}\times\nn
 \ee
 \be\small \left(\begin{array}{ccccccc}
(M+1)^2 & 0 & 0 & 0 & -(M^2-1) & 0 & 0 \\ 0 & (M-1)^2 & 0 & (M^2-1)
& 0 & 0 & 0
\\ 0 & 0 & -2(M^2+1) & 0 & 0 & 0 & \2(M^2-1) \\ 0 &
-(M^2-1) & 0 & -(M+1)^2 & 0 & 0 & 0 \\
(M^2-1) & 0 & 0 & 0 & -(M-1)^2 & 0 & 0 \\ 0 & 0 & 0 & 0 & 0 &
2(M^2+1) & -\2(M^2-1) \\
0 & 0 & -\2(M^2-1) & 0 & 0 & \2(M^2-1) & 0
\end{array}\right)\,.
 \ee
This last charge matrix may be transformed to $\A_S$ by the action
of transformations generated by the isotropy subalgebra
(\ref{isos}). The simplest such transformation
 \be
P_{BS1} = \exp[\beta(q-t)]\,, \quad \beta = -\ln(M)/2
 \ee
leads to the $G_2$ transformation $P_{BS} = P_{BS0}P_{BS1}$
transforming the BTZ black ring into the Schwarzschild black string
 \be
P_{BS} = \frac1{2}\left(\begin{array}{ccccccc} \2M^{-1/2} & 0 & 0 &
0 & -\2M^{-1/2} & 0 & 0 \\ 0 & \2M^{-1/2} & 0 & \2M^{-1/2} & 0 & 0 &
0 \\ 0 & 0 & M & 0 & 0 & M &
\2M \\ 0 & -\2M^{1/2} & 0 & \2M^{1/2} & 0 & 0 & 0 \\
\2M^{1/2} & 0 & 0 & 0 & \2M^{1/2} & 0 & 0 \\
0 & 0 & M^{-1} & 0 & 0 & M^{-1} & -\2M^{-1} \\ 0 & 0 & -\2 & 0 & 0 &
\2 & 0
\end{array}\right)\,.
 \ee
This transformation is not unique, as it can be right-factored by
any transformation generated by the element $r^1-p_1$ of $h_S$,
which commutes with $\A_S = {m_0}^0$.

Conversely, the BTZ black ring belongs to the continuous family of
magnetostatic solutions $M_{\alpha}$ generated from the
Schwarzschild black string by the transformations
 \be
P_{\alpha} = \exp[-\beta(q - t)] \exp[-\alpha(q + t)], \quad \beta =
-\ln(M)/2\,.
 \ee
The non-vanishing scalar potentials
\begin{equation}
\lambda_{00} = - \frac{r - m}{\Sigma}, \quad \lambda_{11} = \frac{r
+ m}{\Sigma}, \quad \nu = -\frac{r \sin(2\alpha) + \frac{a^2 -
m^2}{2a} \cos(2\alpha)}{\Sigma},
\end{equation}
where
\begin{equation}
\Sigma = r \cos(2\alpha) - \frac{a^2 - m^2}{2a} \sin(2\alpha) +
\frac{a^2 + m^2}{2a}\,,
\end{equation}
lead to the five-dimensional solution
\begin{eqnarray}\lb{uprn}
\rd s\5^2 &=& - \frac{r - m}{\Sigma} \rd t^2 + \frac{r + m}{\Sigma}
\rd z^2 + \Sigma^2 \left( \frac{\rd r^2}{r^2 - m^2} + \rd\theta^2 +
\sin^2\theta\,\rd\varphi^2 \right), \nn
\\
A\5 &=& \sqrt3 \left( \frac{a^2 + m^2}{2a} \sin(2\alpha) - \frac{a^2
- m^2}{2a} \right)\,\cos\theta\,\rd\varphi\,.
\end{eqnarray}

This general solution, which includes the Schwarzschild black string
($\cos(2\alpha) = 1, a = m$) and the non-rotating BTZ black ring
($\sin(2\alpha) = -1$) as special cases, can be shown to be an
uplift of the magnetic Reissner-Nordstr\"om solution of
four-dimensional Einstein-Maxwell theory (EM4). Any solution of EM4
can be lifted to a solution of MSG5 given by
\begin{eqnarray}\lb{canup}
\rd s\5^2 &=& \rd s\4^2 + (\rd z + C_\mu dx^\mu)^2\,, \nn\\
A\5 &=& \3A\,\4\,, \qquad \rd C = \star\rd A\4 \,.
\end{eqnarray}
After reduction of (\ref{canup}) to three dimensions, only four
(e.g. $\lambda_{00}$, $\omega_0$, $\psi_0$ and $\nu$) of the eight
scalar potentials are independent, the other four being related to
these by the constraints
 \be
\lambda_{11} = 1\,, \quad \psi_1 = 0\,, \quad \lambda_{01} = \nu\,,
\quad \omega_1 = - \psi_0\,.
 \ee
The solution (\ref{uprn}) does not satisfy these constraints as
written. However these constraints are satisfied in the transformed
coordinate system $(x^0,x^1) = (\tau,\psi)$, with
 \ba
t &=& \chi\big[\sin(\gamma-\alpha)\,\psi - \cos(\gamma-\alpha)\,\tau\big]\,,\\
z &=& \chi\big[\cos(\gamma+\alpha)\,\psi -
\sin(\gamma+\alpha)\,\tau\big]\,,
 \ea
where
 \be
\tan\gamma = \frac{a-m}{a+m}\,, \quad \chi^2 = \frac1{\cos(2\gamma)}
= \frac{a^2+m^2}{2am}\,.
 \ee
After this coordinate transformation, the solution (\ref{uprn}) can
thus be reduced to the four-dimensional magnetic
Reissner-Nordstr\"om solution
\begin{eqnarray}
\rd s\4^2 &=& - \frac{r^2 - m^2}{\Sigma^2} \rd\tau^2 +
\frac{\Sigma^2}{r^2 - m^2}\,\rd r^2 + \Sigma^2\big(\rd\theta^2 +
\sin^2\theta\,\rd\varphi^2 \big), \nn
\\
A\4 &=& \left( \frac{a^2 + m^2}{2a} \sin(2\alpha) - \frac{a^2 -
m^2}{2a} \right)\,\cos\theta\,\rd\varphi\,,
\end{eqnarray}
parameterized in a way which includes the magnetic Bertotti-Robinson
solution.

\setcounter{equation}{0}
\section{From G\"odel to Euclidean Schwarzschild}

The G\"odel solution for minimal supergravity is given by (\ref{gsym}). To
present it in matrix form, it is convenient to introduce
the dimensionless constant $b = 2m/g = \3m/e$, and relabel the
coordinates $t \to 2t$, $y \to \cos\theta$, $ \varphi \to b^2z/m$,  $\psi
\to \varphi$, leading to\footnote{Note that
in passing from (\ref{gsym}) to (\ref{bgods}) we have changed the parity
of the order of appearance of the five-dimensional coordinates, and so
to conform with our convention for the antisymmetric symbol have changed
a sign in $A\5$.}
 \ba\lb{bgods}
\rd s\5^2 &=& -4\left[\rd t - \frac{m}b\,\cos\theta\,\rd\varphi\right]^2 +
\frac{b^2(x^2-1)}{2}\,\rd z^2 + \frac{m^2}{2b^2}
\left(\frac{\rd x^2}{x^2-1} + \rd\theta^2 + \sin^2\theta\,
\rd\varphi^2\right)\,, \nn\\
A\5 &=& 3\left[\rd t - \frac{m}b\,\cos\theta\,\rd\varphi\right] +
\3\,bx\,\rd z\,.
 \ea
This may be toroidally reduced to three dimensions according to the ansatz
(\ref{st5}), leading to the three-dimensional reduced metric (\ref{3red})
and to the metric fields
 \be
\lambda = \mbox{\rm diag}\left(-4,\,\frac{b^2(x^2-1)}2\right) \,,
\quad a_{\varphi} = \left(-\frac{m}b\cos\theta,\, 0\right) \,, \quad \tau =
2b^2(x^2-1)\,,
 \ee
and the electromagnetic fields
 \be
\psi = \left(\3,\,bx\right)\,, \quad
A_{\varphi} = -\frac{\3m}b\,\cos\theta\,,
 \ee
from which one derives the dualized potentials
 \be
\omega = \left(4bx,\, 2\3 b^2x^2\right)\,, \quad \nu = \3 bx
 \ee
(up to irrelevant integration constants). The computation of the
matrix elements leads to a coset representative $M_G(x)$ of the form
(\ref{geo}), with $\sigma = \ln f(x)$, and
 \be\lb{etag}
\eta_G = \frac14\left(\begin{array}{ccccccc} -3 & 0 & 0 & -1 & 0 & 0
& -\6 \\ 0 & 0 & 2\3 & 0 & 2 & 0 & 0 \\ 0 & 2\3 & 0 & 0 & 0
& -2 & 0 \\ -1 & 0 & 0 & -3 & 0 & 0 & \6 \\ 0 & 2 & 0 & 0 & 0 &
2\3 & 0 \\ 0 & 0 & -2 & 0 & 2\3 & 0 & 0 \\ -\6 & 0 & 0 &
\6 & 0 & 0 & 2
\end{array}\right)\,,
 \ee
 \be\lb{ag}
\A_G = \frac12\left(\begin{array}{ccccccc} 0 & 0 & b^{-1} & 0 & 0 &
b & 0 \\ 0 & 0 & 0 & 0 & 0 & 0 & -\2b^{-1} \\ b & 0 & 0 & -b & 0 & 0
& 0 \\ 0 & 0 & -b^{-1} & 0 & 0 & -b & 0 \\ 0 & 0 & 0
& 0 & 0 & 0 & -\2b \\ b^{-1} & 0 & 0 & -b^{-1} & 0 & 0 & 0 \\ 0 & -\2b &
0 & 0 & -\2b^{-1} & 0 & 0
\end{array}\right)\,.
 \ee

We show in Appendix B that, although the three-dimensional reduced
metric is the same, this solution cannot be $G_2$-transformed to
theSchwarzschild black string (\ref{blackS}). This means that the
target space $G_{2(+2)}/((SL(2,R)\times SL(2,R))$ admits at least
two disjoint components, a black string sector generated from the
Schwarzschild black string, and also containing the magnetic
Bertotti-Robinson solution (\ref{bbtz}), as well as black hole
solutions \cite{newdil2}; and a second component containing  the
3-G\"odel solution. We now show that this second component is the
one generated from the Euclidean Scwarzschild string
(\ref{euclS})\footnote{The two solutions of five-dimensional vacuum
gravity (\ref{blackS}) and (\ref{euclS}) are related by analytic
continuation $t \to i z$, $ z \to -it$, so that the two sectors of
$SL(3,R)/SO(2,1)$ generated from these will be related by the same
analytic continuation. But this cannot be extended to the case of
minimal supergravity $G_{2(+2)}/((SL(2,R)\times SL(2,R))$, because
such an analytic continuation would lead to imaginary electric
potentials $\psi_a$.}. Presumably the five-dimensional G\"odel black
holes of \cite{wu} would belong to the first component, however this
remains to be checked.

It is actually very easy to generate from the non-asymptotically
flat G\"odel solution an asymptotically flat solution. $M_G(x)$
leads to a non-asymptotically flat metric because
$\eta_G={M}_{G33}(\infty) = -\tau^{-1}(\infty) = 0$. A generic $G_2$
transformation will lead to ${M'}_{33}(\infty) \neq 0$
(asymptotically flat metric) and with some luck negative (Lorentzian
metric).

An example is the transformation
 \be\lb{p0}
P_0 = \exp{\left[\frac\pi2(\ell_0+n^0)\right]} =
\left(\begin{array}{ccccccc} 0 & 0 & 1 & 0 & 0 & 0 & 0 \\ 0 & 1 & 0
& 0 & 0 & 0 & 0 \\ -1 & 0 & 0 & 0 & 0 & 0 & 0 \\ 0 & 0 & 0 & 0 & 0 & 1 & 0 \\
0 & 0 & 0 & 0 & 1 & 0 & 0 \\ 0 & 0 & 0 & -1 & 0 & 0 & 0 \\ 0 & 0 & 0
& 0 & 0 & 0 & 1 \end{array}\right)\,.
 \ee
Acting on $\eta_G$, this leads to the asymptotic matrix
 \be
\eta'_{0} = P_0^T\eta_GP_0 = \frac14\left(\begin{array}{ccccccc} 0 &
-2\sqrt3 & 0 & -2 & 0 & 0 & 0 \\ -2\sqrt3 & 0 & 0
& 0 & 2 & 0 & 0 \\ 0 & 0 & -3 & 0 & 0 & -1 & -\sqrt6 \\
-2 & 0 & 0 & 0 & -2\sqrt3 & 0 & 0 \\
0 & 2 & 0 & -2\sqrt3 & 0 & 0 & 0 \\ 0 & 0 & -1 & 0 & 0 & -3 & \sqrt6 \\
0 & 0 & -\sqrt6 & 0 & 0 & \sqrt6 & 2 \end{array}\right)\,,
 \ee
corresponding to an asymptotically Lorentzian (up to a coordinate
transformation) $\lambda'_{ab}(\infty)\,\rd x^a\rd x^b = -(4/\3)\,\rd
x^0\rd x^1$ (and thus to an asymptotically Lorentzian five-metric),
with $\omega'_a(\infty) = 0$, $\psi'_a(\infty) = 0$, $\nu'(\infty) =
-1/\3$. A gauge transformation $Q$ (linear transformation in $(\rd
x^a, \;\rd x^b)$ together with a translation of $\nu$) will then
transform $\eta'_{0}$ to the vacuum Lorentzian form
 \be
\eta' = Q^T\eta'_0Q = \eta_S\,.
 \ee
The corresponding full coset matrix will be
 \be
M'( r) = \eta_S\,\e^{\A_0'\sigma( r)}\,,
 \ee
with
 \be
\A_0' = Q^{-1}P_0^{-1}\A_GP_0Q\,.
 \ee

The computation gives ($\beta = 3^{1/4}$)
 \be
P_0Q = \left(\begin{array}{ccccccc} 0 & 0 & 2\beta^{-2} & 0 & 0 &
\ds\frac{\beta^{-2}}2 & -\2\beta^{-2} \\
\ds\frac\beta2 & \ds\frac\beta2 & 0 & 0 & 0 & 0 & 0 \\
-\ds\frac\beta2 & \ds\frac\beta2 & 0 & 0 & 0 & 0 & 0 \\ 0 & 0 & 0
& 0 & 0 & \ds\frac{\beta^2}2 & 0 \\ -\ds\frac{\beta^{-1}}2 &
\ds\frac{\beta^{-1}}2 & 0 & \beta^{-1} & \beta^{-1} & 0 & 0
\\ -\ds\frac{\beta^{-1}}2 & -\ds\frac{\beta^{-1}}2 & 0 & -\beta^{-1} &
\beta^{-1} & 0 & 0 \\ 0 & 0 & 0 & 0 & 0 & -\ds\frac1\2 & 1
\end{array}\right)\,,
 \ee

 \ba
A'_0 &=& \beta^{-3}[b(n^0-\ell_0) + b(n^1+\ell_1) - \gamma(p_0-r^0)
- \delta(p_1+r^1)] \nn\\ && (\gamma = (\3b^{-1}-b)/2\,,\; \delta
= (\3b^{-1}+b)/2)\,.
 \ea
This special charge matrix includes a NUT charge, proportional to
the coefficient $-b$, a Kaluza-Klein magnetic charge,
proportional to $b$, and two electric charges (the fluxes of $F_{0
r}$ and $F_{1 r}$), proportional to $-\gamma$ and $-\delta$.

This may be diagonalized to
 \be
A'_1 = P_1^{-1}A'_0P_1 = {\rm diag}(0,\;1,\;-1,\;0,\;-1,\;1,\;0)
 \ee
through the action of transformations
generated by the isotropy subalgebra (\ref{isos}):
 \be
P_1 =  \e^{\alpha_1({m_1}^0+{m_0}^1)}\e^{(\pi/4)(p_1-r^1)}
\e^{-(\pi/4)(\ell_0+n^0)}\e^{-2\alpha_2({m_1}^0+{m_0}^1)}
 \ee
with $\e^{\alpha_1} = \beta^{-1}b^{-1}$, $\e^{\alpha_2} = \beta^{-1}$.
Putting everything together, we have transformed
$M_G( r)$ by the transformation
 \be
P = P_0QP_1
 \ee
to the diagonal form
 \be
M_{ES} = \mbox{\rm diag} (-1,\; f,\; -f^{-1},\; -1,\; f^{-1},\;
-f,\; 1)\,,
 \ee
corresponding to the Euclidean Schwarzschild string (\ref{euclS})
with $A\5 = 0$.

\setcounter{equation}{0}
\section{Generating AF soliton solutions}
\subsection{A continuous family of NUTty soliton solutions}

A generic $G_2$ transformation acting on $M_G(x)$ can lead either
to $\tau^{-1}(\infty) > 0$, corresponding to a five-dimensional
metric with signature $(-++++)$ (asymptotically flat soliton
strings, as in the preceding section), or to $\tau^{-1}(\infty) <
0$, which could correspond to a five-dimensional metric with either
the signature $(-----)$ (asymptotically flat five-dimensional
anti-instantons), or the signature $(---++)$. A continuous family
containing both soliton strings and anti-instantons can be generated
from $M_G(x)$ by the $SL(3,R)$ transformation
 \be\lb{cont} P_{\alpha} = \exp{\left[\alpha(\ell_1+n^1)\right]} =
\left(\begin{array}{ccccccc} 1 & 0 & 0 & 0 & 0 & 0 & 0 \\ 0 & c & s
& 0 & 0 & 0 & 0 \\ 0 & -s & c & 0 & 0 & 0 & 0 \\ 0 & 0 & 0 & 1 & 0 & 0 & 0 \\
0 & 0 & 0 & 0 & c & s & 0 \\ 0 & 0 & 0 & 0 & -s & c & 0 \\ 0 & 0 & 0
& 0 & 0 & 0 & 1 \end{array}\right)\,,
 \ee
with $s \equiv \sin\alpha$, $c \equiv \cos\alpha$. This leads to the
transformed scalar potentials
\begin{eqnarray}
\tau' &=& - \frac{2(x^2 - 1)}{\sqrt3 s_2 \, x^2 - c_2 \, \sigma_+ +
\sigma_-}, \nn
\\
\nu' &=& - \frac{( sb + \sqrt3 c {b}^{-1} ) x}{ \sqrt3 s_2 \, x^2 -
 c_2 \, \sigma_+ + \sigma_-}, \nn
\\
\psi' &=& \left( \begin{array}{c} -\ds\frac{s_2 \, x^2 + \sqrt3 (c_2
\, \sigma_+ - \sigma_-)}{\sqrt3 s_2 \, x^2 - c_2 \, \sigma_+ +
\sigma_-}
\\
-\ds\frac{( c {b}^{-1} - \sqrt3 s  b ) x}{\sqrt3 s_2 \, x^2 - c_2 \,
\sigma_+ + \sigma_-}
\end{array} \right), \nn
\\
\omega' &=& \left( \begin{array}{c} \ds\frac{4 c { b}^{-1} (2 s^2 {
b}^2 x^2 + c_2 \, \sigma_+ - \sigma_-) x}{(\sqrt3 s_2 \, x^2 - c_2
\, \sigma_+ + \sigma_- )^2}
\\
\ds\frac{-3 c_2 \, s_2 \, x^4 - s_2 \, x^2 + 2 \sqrt3 c^2 ( 2 c_2 \,
\sigma_+ - { b}^2 ) x^2 + s_2 \, \sigma_+ (c_2 \, \sigma_+ -
\sigma_-)}{\left(\sqrt3 s_2 \, x^2 - c_2 \, \sigma_+ + \sigma_-
\right)^2}
\end{array} \right), \nn
\\
\lambda'_{00} &=& - \ds\frac{4 \left[ s_2^2 \, x^4 + (c_2 \,
\sigma_+ - \sigma_-)^2 \right]}{\left(\sqrt3 s_2 \, x^2 - c_2 \,
\sigma_+ + \sigma_-
\right)^2}\,, \nonumber\\
\lambda'_{01} &=& - \ds\frac{4 s  b \left( 2 c^2 { b}^{-2} x^2 - c_2
\, \sigma_+ + \sigma_- \right) x}{\left(\sqrt3 s_2 \, x^2 - c_2 \,
\sigma_+ +
\sigma_- \right)^2}\,, \nonumber\\
\lambda'_{11} &=& -\ds\frac{3 \sqrt3 s_2 x^2 (x^2 - 1) + (9 s^2 {
b}^2 - c^2 { b}^{-2}) x^2 + c_2 \, \sigma_+ -
\sigma_-}{2\left(\sqrt3 s_2 \, x^2 - c_2 \, \sigma_+ + \sigma_-
\right)^2},
\end{eqnarray}
where
$$ s_2 \equiv \sin2\alpha\,, \quad c_2 \equiv \cos2\alpha\,, \quad
\sigma_{\pm} \equiv \frac{b^2 \pm b^{-2}}2\,. $$

From the expression of $\tau'$, one sees that the corresponding
solution is a soliton string for $s_2 < 0$, and an anti-instanton
for $s_2
> 0$. The non-asymptotically flat divides between the two ($s_2 = 0$)
correspond to the G\"odel string (\ref{gsym}) for $\sin\alpha = 0$
(with both signs of $\cos\alpha$ possible), and to the anti-G\"odel
solution for $\cos\alpha = 0$, see below.

Inverse dualization, carried out according to
(\ref{dualmu})-(\ref{dualom}), with $\varepsilon = -{\rm sign}(s_2)$
leads to
\begin{eqnarray}
a_\varphi' &=& \varepsilon m\,\left( \begin{array}{c}   -c { b}^{-1}
\cos\theta
\\ 0
\end{array} \right),
\\
A_\varphi' &=& \varepsilon bm\,\frac{s_2 (c { b}^{-2} + \sqrt3 s )
x^2 - (s - \sqrt3 c { b}^{-2} ) (c_2 \, \sigma_+ - \sigma_-)}{\sqrt3
s_2 \, x^2 - c_2 \, \sigma_+ + \sigma_-} \cos\theta \nn \,.
\end{eqnarray}
The resulting five-dimensional metric and gauge can be put in a
simple form by defining the real parameter $\beta^2 = -\3s_2/2$, and
making the coordinate redefinitions
 \ba
t &\to& \frac{\3}2\,t\,, \quad x \to \frac{r}{\mu}\,, \quad  z \to
\frac2{\3}\,\beta\,z \qquad (\beta^2 > 0)\,, \nn\\
t &\to& \frac{\3}2\,t\,, \quad r \to i\frac{r}{\mu}\,, \quad  z \to
-\frac2{\3}\,i\beta\,z \qquad (\beta^2 < 0)\,,
 \ea
leading to
 \ba\lb{solprime}
\rd s\5^{'2} &=& - \frac{r^4+3\nu^4}{(r^2-\nu^2)^2}\left[\rd t +
2\varepsilon N\cos\theta\,\rd\varphi +
\frac{2(Nr^2+P\nu^2)r}{r^4+3\nu^4}\,\rd z \right]^2 \nn\\  && +
\varepsilon(r^2-\nu^2)\left[ \frac{r^2-\mu^2}{r^4+3\nu^4}\,\rd z^2 +
\frac{\rd r^2}{r^2-\mu^2} + \rd\theta^2 +
\sin^2\theta\,\rd\varphi^2\right]\,, \\
A'\5 &=& -\frac{\3}2\left[\frac{r^2+3\nu^2}{r^2-\nu^2}(\rd t +
2\varepsilon N\cos\theta\,\rd\varphi)+ \frac{2(N+P)r}{r^2-\nu^2}\,\rd z
- 2\varepsilon P\cos\theta\,\rd\varphi\right]\,. \nn
 \ea
This solution, with signature $(-++++)$ for $\varepsilon > 0$, or
$(-----)$ for $\varepsilon < 0$, depends on two real parameters $N$
(NUT charge) and $P$ (the magnetic charge is $N-P$), with
 \ba
N &=& -\frac{mc}{\3\,b}\,, \quad P = bms\,, \nn\\
\mu^2 &=& m^2|\beta^2| = 3\varepsilon NP\,, \nn\\
\nu^2 &=& -\varepsilon m^2\frac{(c_2\sigma_+-\sigma_-)}2 =
\varepsilon\frac{P^2-3N^2}2\,.
 \ea

For the exceptional value $\alpha = \varepsilon'\pi/2$ ($c = 0$, $s
= \varepsilon'$, $s_2 = 0$, $c_2 = -1$, $-c_2\sigma_++\sigma_- =
b^{2}$), $\tau'$ is negative in the sector $x^2>1$ and, after
inverse dualisation according to (\ref{dualmu})-(\ref{dualom}) and
time rescaling $t\to t/2$, the five-dimensional solution reduces to
 \ba\lb{godcont}
\rd s\5^{'2} &=& -\left(\rd t - g'x
\,\rd\psi\right)^2 - \frac{g'^2}8\left[\frac{\rd x^2}{x^2-1}
+ (x^2-1)\rd\psi^2 + \frac{\rd y^2}{1-y^2} +
(1-y^2)\rd\varphi^2\right] \nn\\ A\5' &=& \frac32\left(\rd t -
g'x\,\rd\psi\right) + \frac{\3}2\,g'y\,\rd\varphi\,,
 \ea
where we have put $y = \cos\theta$, $\psi = z/b^2m$, and
$g'=-2\varepsilon'bm$. This is recognized as the anti-G\"odel
solution in the symmetric form (\ref{gsym}) with $g\to g'$ and the
coordinate relabellings $x \leftrightarrow y$.

The metric (\ref{solprime}) has a bolt at $r^2 = \mu^2$, where it is
regular (if $\nu^2 \neq \mu^2$) provided the coordinate $z$ is
periodically identified with period $\pi\3(P^2+3N^2)/\mu$, and is
singular at $r^2 = \nu^2$, unless $\nu^2 < 0$. It follows that (with
this periodic identification of the coordinate $z$) this
two-parameter solution is for $\varepsilon=+1$ a regular soliton
ring provided the ratio of the two parameters lies in the range
 \be
0 < \frac{P}{N} < 3 + 2\sqrt3\,.
 \ee
The  bolt at $r^2 = \mu^2$ is extreme for $\mu^2 = 0$. The
near-extreme, near-bolt regime corresponds to
 \be
\mu = |\beta|m\,, \quad r =  |\beta|\ol{r} \qquad (\beta \to 0)\,.
 \ee
This can be achieved in two ways. Either $N$ is held fixed, and $P$
goes to zero ($\nu^2 = -\varepsilon\,3N^2/2$) with $|\beta^2|$,
leading (up to coordinate rescalings) to the G\"odel solution in its
original form (\ref{bgods}). Or $P$ is held fixed, and $N$ goes to
zero ($\nu^2 = \varepsilon P^2/2$) with $|\beta^2|$, leading (again
up to coordinate rescalings) to the anti-G\"odel solution
(\ref{godcont}).

\subsection{A class of NUTless, massive solitons}

The asymptotically locally flat solution (\ref{solprime}) presents two
defects: 1) it is massless; 2) its NUT singularity prevents it from
being truly asymptotically flat. Both defects can be cured by acting
on this solution with the NUT-to-mass transformation. This $SL(2,R)$
transformation, which transforms the massless Schwarzschild-NUT solution
of vacuum gravity into the Schwarzschild solution, is
 \be
P_{MN}=\exp[(\pi/4)(n_0+\ell^0)]\,.
 \ee
The action of this transformation on the charge matrix  associated
with the solution (\ref{solprime}),
\begin{equation}
\A'= -\frac1{2\mu}\left(
\begin{array}{ccccccc}
 0 & 2 N & 2 N & 0 & P-N & N+P & 0 \\
 -2 N & 0 & 0 & N-P & 0 & 0 & -\sqrt{2} (N+P) \\
 2 N & 0 & 0 & -(N+P) & 0 & 0 & \sqrt{2} (N-P) \\
 0 & P-N & -(N+P) & 0 & 2 N & -2 N & 0 \\
 N-P & 0 & 0 & -2 N & 0 & 0 & -\sqrt{2} (N+P) \\
 N+P & 0 & 0 & -2 N & 0 & 0 & \sqrt{2} (P-N) \\
 0 & -\sqrt{2} (N+P) & \sqrt{2} (P-N) & 0 & -\sqrt{2} (N+P) & \sqrt{2} (N-P) & 0
\end{array}\right)\,.
\end{equation}
leads to the transformed charge matrix $\A'' = P_{MN}^{-1}A'P_{MN}$:
\begin{equation}
\A'' = -\frac1{2\mu}\left(
\begin{array}{ccccccc}
-2 N & \sqrt{2} N & 0 & 0 & \frac{P-N}{\sqrt{2}} & N+P & P-N \\
 -\sqrt{2} N & 0 & -\sqrt{2} N & \frac{N-P}{\sqrt{2}} & 0 & \frac{N-P}{\sqrt{2}} & -\sqrt{2} (N+P) \\
 0 & \sqrt{2} N & 2 N & -(N+P) & \frac{P-N}{\sqrt{2}} & 0 & N-P \\
 0 & \frac{P-N}{\sqrt{2}} & -(N+P) & 2 N & \sqrt{2} N & 0 & N-P \\
 \frac{N-P}{\sqrt{2}} & 0 & \frac{N-P}{\sqrt{2}} & -\sqrt{2} N & 0 & -\sqrt{2} N & -\sqrt{2} (N+P) \\
 N+P & \frac{P-N}{\sqrt{2}} & 0 & 0 & \sqrt{2} N & -2 N & P-N \\
 N-P & -\sqrt{2} (N+P) & P-N & P-N & -\sqrt{2} (N+P) & N-P & 0
\end{array}
\right)
\end{equation}
The resulting coset representative $M''= \eta_se^{\A''\sigma}$ leads
to the solution
 \ba\lb{solsec}
\rd s_5^{''2} &=& \lambda_{00}\left[\rd
t+\frac{\lambda_{01}}{\lambda_{00}}(\rd z -
\sqrt2N\cos\theta\,\rd\varphi)\right]^2 -
\frac{\tau}{\lambda_{00}}(\rd z - \sqrt2N\cos\theta \rd\varphi)^2 +
\frac{\rd r^2}{\tau} +
\frac{r^2-\mu^2}\tau(\rd\theta^2+\sin^2\theta\rd\varphi^2)\,, \nn\\
A''_5 &=&
\3\left[\frac{(N-P)r+3N^2-P^2}{\sqrt2(r-\alpha)(r-\beta)}\,\rd t -
\frac{N+P}{r-\beta}(\rd z - \sqrt2N\cos\theta\,\rd\varphi) -
\frac{N-P}{\sqrt2} \,\cos\theta \rd\varphi \right]\,,
 \ea
with
 \ba
\tau &=& \frac{r^2-\mu^2}{(r-\alpha)(r-\beta)}\,, \quad \lambda_{11} =
\frac{(r-\alpha)(r-2\beta+\alpha)}{(r-\beta)^2}\,, \nonumber\\
\lambda_{01}&=&\frac{(3 N^2+4NP -3 P^2-4 N r)}{2\sqrt2(r-\beta)^2}\,, \quad
\lambda_{00} = \frac{\lambda_{01}^2 - \tau}{\lambda_{11}}\,,
 \ea
where
\begin{equation}
\alpha = -\frac{3N+P}2\,, \quad \beta = \frac{-N+P}2\,, \quad \mu^2
= 3NP\,.
\end{equation}

Assuming $\mu^2>0$, the metric (\ref{solsec}) has the Minkowskian
signature ($\tau$ is positive) for $r>\mu$ if
\begin{equation}
\mu-\alpha = \frac{(\mu+3N)^2}{6N} >0\,, \quad \mu-\beta =
\frac{3N^2+6\mu N -\mu^2}{6N} >0\,.
\end{equation}
The first inequality is ensured if $N>0$, the second is then ensured
if
\begin{equation}\label{bound}
\frac{N}{\mu} > \frac2{\sqrt3}-1 = 0.154701\,.
\end{equation}
Near the bolt $r=\mu$, $\lambda_{00} \simeq
-\lambda_{01}^2/\lambda_{11}$. One can check that
$\lambda_{11}(\mu)$ is negative definite, implying
$\lambda_{00}(\mu)$ negative definite. One can also check that
$\lambda_{00}$ is finite at the zero $r=2\beta-\alpha$ of
$\lambda_{11}$ (and thus is negative definite in the range $r \ge
\mu$), implying that $\lambda_{01}^2-\tau$ can be factored by
$r-2\beta+\alpha$, so that the expression of $\lambda_{00}$ can be
simplified, but is still somewhat cumbersome.

For the absence of conical singularity, the coordinate $z$ must be
periodically identified with period
\begin{equation}
T =
(-\lambda_{00}(\mu))^{1/2}\frac{(\mu-\alpha)(\mu-\beta)}{\mu}2\pi =
\sqrt{\frac23}\frac{(3N+P+2\mu)(3N+3P-2\mu)}{4\mu}2\pi\,.
\end{equation}
The Misner string singularity is absent if this period is equal to
$\sqrt2N\times4\pi$, leading to the quartic equation
\begin{equation}
27N^4-24\sqrt3N^3\mu+4N\mu^3+\mu^4=0\,.
\end{equation}
This equation has the two real solutions
 \begin{equation}
N = 0.460230\mu\,, \quad N = 1.45795\mu\,,
 \end{equation}
both satisfying the bound (\ref{bound}). For these values of the
ratio $N/\mu$, the solution ($ds^{''2}_5$, $A''_5$) is a soliton
ring with mass $N$.

\setcounter{equation}{0}
\section{Multicenter solutions}
Null-geodesic solutions are of
the form (\ref{geo}), with the charge matrix $\A$ constrained by
the charge balance condition \cite{spat,bps}
 \be
{\rm Tr}(\A^2) = 0\,.
 \ee
Null geodesics lead to a Ricci-flat, hence
flat, reduced 3-space of metric $h_{ij}$ \cite{spat,bps}. In that
case, the Laplacian $\nabla_h^2$ becomes a linear operator, so that
an arbitrary number of harmonic functions may be superposed, leading
to a multicenter solution
 \be\lb{sigmult}
\sigma(\vec{x}) = \epsilon + \sum_i\frac{a_i}{|\vec{x}-\vec{x_i}|}\,.
 \ee

It is easy to promote the special solutions presented in Sect. 2 to
multicenter (null geodesic) solutions, provided that, after toroidal
reduction relative to $\partial_t$ and $\partial_{z}$, the reduced
metric is flat. This is the case for the self-dual solution
(\ref{bsd}), as well as its $c=0$ limit, the extreme BTZ solution
(\ref{bbtz}) with $J^2 = 4M^2a^2$, and for the G\"odel solution
(\ref{bgods}) in the extreme case $m^2=0$ with $b/m=2/g$ fixed.

\subsection{Self-dual solutions}
We first consider the self-dual solution
(\ref{bsd}) which contains for $c=0$ the
extreme BTZ solution. This solution
can be generalized by replacing the harmonic function $a/r$ by an
arbitrary harmonic function $\sigma(\vec{x})$,
 \ba\lb{multisd}
\rd s\5^2 &=& \sigma^{-1}\,\rd u\,\rd v + \left(M -
\frac{3c^2}{4\pm1}\, \sigma^{\pm4} \right)
\rd u^2 + \sigma^2\,\rd\vec{x}^2\,, \nn\\
A\5 &=& \3\left[c\,\sigma^{\pm2}\,\rd u \pm \,A_3\right]
\qquad (\nabla\wedge A_3 = \nabla\sigma)\,,
 \ea
with $u = z-t$, $v = z+t$ \footnote{Again, we have changed a sign in
$A\5$ because our coordinate transformation implies
$\epsilon_{uv}=-\epsilon_{tz}$.}. The linear superposition
(\ref{sigmult}) leads to multicenter solutions of MSG5, which are
asymptotic to the one-center solution (\ref{bsd}) for $\epsilon=0$,
and asymptotically Minkowskian (up to a gauge transformation) for
$\epsilon=1$. As shown in \cite{reds2}, the one-center
asymptotically Minkowskian solution (\ref{multisd}) is an extreme
black string for the lower sign, while the spacetime is geodesically
complete for the upper sign.

The scalar potentials associated with (\ref{bsd}) are (with $x = r/a$)
 \be
\lambda = \left(\begin{array}{cc} -x+M_{\pm}(x) & -M_{\pm}(x) \\
-M_{\pm}(x) & x+M_{\pm}(x)
\end{array}\right) \,, \quad \tau = x^2\,, \quad
\psi = c\,x^{\mp2}(1,\;-1)\,,
 \ee
with
 \be
M_{\pm}(x) = M - \frac{3c^2}{4\pm1}\,x^{\mp4}\,,
 \ee
and
 \be
\omega = \left\vert\begin{array}{c} -3c\,x^{-1}(1,\;-1) \\ -c
\,x^3(1,\;-1) \end{array} \right. \,, \quad \nu = \pm x\,.
 \ee

The character of the null-geodesic solutions depends crucially on
the choice of the sign $\pm$. For the lower sign, the representative
matrix, where we have replaced $x$ by $\sigma^{-1}$ (assuming
$\epsilon=0$ in (\ref{sigmult})) is
 \be\lb{Mmultisddown}
M = \left(\begin{array}{ccccccc} 0 & 0 & 0  & -M\sigma & 1-M\sigma & 0 & 0 \\
0 & 0 & 0  & 1+M\sigma & M\sigma & 0 & 0 \\ 0 & 0 & -\sigma^2 & c & c & -1 &
-\2\sigma \\ -M\sigma & 1+M\sigma & c & -\sigma-M\sigma^2 & -M\sigma^2 & 0 & 0 \\
1-M\sigma & M\sigma & c & -M\sigma^2 & \sigma-M\sigma^2 & 0 & 0 \\ 0 & 0 & -1 &
0 & 0 & 0 & 0 \\  0 & 0 & -\2\sigma & 0 & 0 & 0 & -1
\end{array}\right) \,.
 \ee
The charge matrix
 \be
\A = \left(\begin{array}{ccccccc} -M & M & 0  & 0 & 1 & 0 & 0 \\
-M & M & 0  & -1 & 0 & 0 & 0 \\ 0 & 0 & 0 & 0 & 0 & 0 & 0 \\
0 & 0 & 0 & M & M & 0  & 0 \\  0 & 0 & 0 & -M & -M & 0  & 0 \\
0 & 0 & 0 & 0 & 0 & 0  & \2 \\ 0 & 0  & \2 & 0 & 0 & 0 & 0
\end{array}\right)
 \ee
does not depend on the parameter $c$ (which enters only the asymptotic matrix $\eta$),
and is such that
 \be\lb{A32}
\A^3 = 0\,, \quad \A^2 \neq 0\,.
 \ee
The solution is presumably a $G_2$ transform of the vacuum
(anti-)self-dual solution given in \cite{spat}, with equal
Kaluza-Klein electric and magnetic charges and a nilpotent charge
matrix obeying (\ref{A32}).

For the upper sign, the representative matrix is of the form
(\ref{geo}), with
 \ba
\eta &=& \left(\begin{array}{ccccccc} 0 & 0 & 0  & 0 & -1 & 0 & 0 \\
0 & 0 & 0  & -1 & 0 & 0 & 0 \\ 0 & 0 & 0 & 0 & 0 & -1 & 0 \\
0 & -1 & 0 & 0 & 0 & 0 & 0 \\  -1 & 0 & 0 & 0 & 0 & 0 & 0 \\
0 & 0 & -1 & 0 & 0 & 0 & 0 \\ 0 & 0 & 0 & 0 & 0 & 0 & -1
\end{array}\right)\,, \\
\A &=& \left(\begin{array}{ccccccc} -M & M & 0  & 0 & -1 & 0 & 0 \\
-M & M & 0  & 1 & 0 & 0 & 0 \\ 12c & -12c & 0 & 0 & 0 & 0 & 0 \\
0 & 0 & 0 & M & M & -12c & 0 \\  0 & 0 & 0 & -M & -M & 12c & 0 \\
0 & 0 & 0 & 0 & 0 & 0  & -\2 \\ 0 & 0  & -\2 & 0 & 0 & 0 & 0
\end{array}\right)\,.
 \ea
For $c\neq0$, this charge matrix is nilpotent of rank six, i.e.
 \be
\A^7 = 0\,, \quad \A^6 \neq 0\,.
 \ee
The corresponding geodesically complete, asymptotically
$AdS_3\times S^2$, multicenter solution has no vacuum counterpart.
In the notations of \cite{KHPV}, it belongs to the orbit ${\cal O}_5$
of $G_{2(2)}$, which also contains the supersymmetric G\"odel black hole \cite{BerkPiol}.

\subsection{G\"odel solutions}
Trading the radial coordinate $x$ of the G\"odel solution
(\ref{bgods}) for $r=mx$, taking the limit $m\to0$ with $g=2m/b$
fixed, and replacing the harmonic function $g/r$ by an arbitrary
harmonic function $\sigma(\vec{x})$ leads to the solution
 \ba
\rd s\5^2 &=& -(2\rd t - A_3)^2 + 2\sigma^{-2}\,\rd z^2 +
\frac{\sigma^2}8\,\rd\vec{x}^2\,, \nn\\
A\5 &=& \frac32(2\rd t - A_3) + 2\sigma^{-1}\,\rd z\,.
 \ea

The corresponding scalar potentials are
 \ba
\lambda &=& \left(
\begin{array}{cc}
 -4 & 0 \\
 0 & 2\sigma ^{-2}
\end{array}
\right)\,, \quad \omega=8\left(
\begin{array}{c}
 \sigma^{-1} \\
 \sqrt{3}\sigma^{-2}
\end{array}
\right)\,, \nn\\
\psi &=& \left(
\begin{array}{c}
 \sqrt{3} \\
 2\sigma^{-1}
\end{array}
\right)\,, \quad \nu=2\sqrt{3}\sigma^{-1}\,.
 \ea
The representative matrix
 \be\lb{Mmultigod}
\left( \begin{array}{ccccccc}
 -3 & 0 & 2\sigma  & -1 & -2 \sqrt{3} \sigma  & 0 & -\sqrt{6} \\
 0 & 0 & 2 \sqrt{3} & 0 & 2 & 0 & 0 \\
 2\sigma  & 2 \sqrt{3} & -2 \sigma^2 & -2\sigma  & 2 \sqrt{3}\sigma ^2 & -2 & 2 \sqrt{6}\sigma  \\
 -1 & 0 & -2\sigma  & -3 & 2 \sqrt{3}\sigma  & 0 & \sqrt{6} \\
 -2 \sqrt{3}\sigma  & 2 & 2 \sqrt{3} \sigma ^2 & 2 \sqrt{3}\sigma  & 2 \sigma ^2 & 2 \sqrt{3} & 2 \sqrt{2}
   \sigma  \\
 0 & 0 & -2 & 0 & 2 \sqrt{3} & 0 & 0 \\
 -\sqrt{6} & 0 & 2 \sqrt{6}\sigma  & \sqrt{6} & 2 \sqrt{2}\sigma  & 0 & 2
\end{array}
\right)
 \ee
is of the form (\ref{geo}), with $\eta$ given by (\ref{etag}), and the charge matrix
 \be
{\cal A}= \frac 12\left(
\begin{array}{ccccccc}
 0 & 0 & -1 & 0 & 0 & 0 & 0 \\
 0 & 0 & 0 & 0 & 0 & 0 & \sqrt{2} \\
 0 & 0 & 0 & 0 & 0 & 0 & 0 \\
 0 & 0 & 1 & 0 & 0 & 0 & 0 \\
 0 & 0 & 0 & 0 & 0 & 0 & 0 \\
 -1 & 0 & 0 & 1 & 0 & 0 & 0 \\
 0 & 0 & 0 & 0 & \sqrt{2} & 0 & 0
\end{array}
\right)\,,
 \ee
which is idempotent of rank two, ${\cal A}^3=0$, $\A^2 \neq 0$. The
question of whether the representative matrices (\ref{Mmultigod})
and (\ref{Mmultisddown}) can be transformed into each other, or
belong to two inequivalent components of the ${\rm Tr}(\A^2) = 0$
sector of solution space, remains open.

\setcounter{equation}{0}
\section{Application to the generation of rotating AF solutions}
Toroidal reduction can also be performed relative to two linearly
independent combinations of the three Killing vectors. Replacing
e.g. $\partial_t$ by a linear combination of $\partial_t$ and
$\partial_{ z}$ simply amounts to changing the values of the
parameters $M$ and $J$, or $m$ and $\omega$. On the other hand,
replacing $\partial_t$ by a linear combination of $\partial_t$ and
$\partial_{\varphi}$ should, as in the four-dimensional
Einstein-Maxwell case \cite{kerr}, lead to rotating solutions.

As mentioned in Sect. 4, any solution of EM4 can be lifted to a
solution (\ref{canup}) of MSG5. Applying this lifting procedure to
the four-dimensional electric Bertotti-Robinson solution, with the
spacetime geometry $AdS_2\times S^2$, one obtains
\cite{spinem5,Clement:2008qx} a five-dimensional electric
Bertotti-Robinson solution with the geometry $AdS_2\times S^3$,
while the four-dimensional magnetic Bertotti-Robinson solution lifts
to the five-dimensional magnetic Bertotti-Robinson solution
(\ref{bbtz}) with $J=0$, with the geometry $AdS_3\times S^2$, and
the continuous family of four-dimensional dyonic Bertotti-Robinson
solutions lifts to five-dimensional solutions with geometries
interpolating between $AdS_2\times S^3$ and $AdS_3\times S^2$.

Thus, the EM4 spin-generating mechanism of \cite{kerr} can be lifted
to the case of MSG5 in several fashions. In all cases, this
generation will proceed in three steps. First, carry out a
transformation $\Pi$ from an asymptotically flat static solution to
the corresponding asymptotically Bertotti-Robinson solution. Second,
perform on this the combined transformation
 \be\lb{rot}
\rd\varphi' = \rd\varphi - \Omega\,\rd t\,, \quad \rd t' =
\alpha^{-1}\rd t\,,
 \ee
which does not modify the leading asymptotically Bertotti-Robinson
behavior, but modifies the three-dimensional reduced metric
$\rd\sigma^2$. For instance, the reduced metric (\ref{3red}) is
transformed into
 \be
\rd\sigma'^2 = \frac{\hat\tau'}{\hat\tau}\left[\rd r^2 +
(r^2-r_0^2)\rd\theta^2\right] +
\alpha^2(r^2-r_0^2)\sin^2\theta\,\rd\varphi^2\,,
 \ee
where $\hat\tau$ and $\hat\tau'$ refer to the untransformed and
transformed Bertotti-Robinson metrics, with
 \be
\hat\tau' = \alpha^2\left[\hat\tau -
\Omega^2\hat\tau^{-1}\hat\lambda_{11}(r^2-r_0^2)\sin^2\theta\right]\,.
 \ee
Third, transform back with $\Pi^{-1}$ to an asymptotically flat
rotating solution. If the input static solution is uncharged, the
output rotating solution will also be uncharged for a suitable value
of the parameter $\alpha$ \cite{kerr}.

If the input static solution is a Tangherlini black hole, this
procedure should lead \cite{spinem5} to a Myers-Perry black hole.
The details have not been spelled out in \cite{spinem5}, but to
obtain a black hole with two independent angular momenta one should
presumably generalize (\ref{rot}) to a combined transformation
 \be
\rd\varphi' = \rd\varphi - \Omega_{\varphi}\,\rd t\,, \quad \rd z' =
\rd z - \Omega_{ z}\,\rd t\,,\quad \rd t' = \alpha^{-1}\rd t\,.
 \ee
The same procedure can be applied to generate a rotating solution
from any static solution of EM5 with Tangherlini asymptotics. The
application to the (singular) static Emparan-Reall black ring (which
has the same asymptotics as a black hole) was carried out in
\cite{spinem5} (using for $\Pi$ the transformation from Tangherlini
to the electric Bertotti-Robinson solution, and the transformation
(\ref{rot})), with inconclusive results.

The same procedure applied to a static black string, using for $\Pi$
the transformation from the Schwarzschild black string to the
magnetic Bertotti-Robinson solution ((\ref{bbtz}) with $J = 0$)
should lead to a rotating black string. Rotating black strings can
also be obtained from rotating black holes by the black hole to
black string transformation of \cite{newdil2}, but it is not clear
whether the two procedures always lead precisely to the same
solutions. Conceively, the resulting solutions might differ by
higher multipole moments. One could also apply the spin-generating
procedure to either a black string or a black hole with a
five-dimensional dyonic Bertotti-Robinson solution as intermediate.

This procedure could also in principle be carried out to generate
spinning soliton strings or five-dimensional anti-instantons, the
magnetic Bertotti-Robinson solution being replaced by the ``rotating
Bertotti-Robinson'' solution equivalent of (\ref{bgods}) obtained by
the coordinate transformation $ t \to z$, $z \to -t$ (a $G_{2(+2)}$
transformation).

\section{Conclusion}
In this paper we have demonstrated the possibility of transforming
non-asymptotically flat solutions into asymptotically flat ones
using sigma-model maps between different classes of geodesic
solutions. This opens a way to construct global black hole solutions
starting with near-horizon solutions as seeds. Though we restrained
ourselves to the special case of five-dimensional minimal
supergravity, this possibility looks general and deserves further
study. We have revealed some general features of $AF \leftrightarrow
NAF$ maps, and provided a particular realization transforming the
Bertotti-Robinson-type solution related to the three-dimensional
G\"odel black hole into new NUTty or NUTless asymptotically flat
soliton ring solutions of MSG5. In the NUTless case, this new ring
is horizonless and contains neither conical, nor Misner string
singularities. Its physical properties and possible applications
await to be investigated.

We have also explored one subtle point in the sigma-model generating
techniques concerning transformations between solutions possessing
the same reduced three-metric and the same asymptotics, which
correspond to geodesics passing through the same point in target
space. Such solutions are defined by the tangent  vectors to
geodesics at this point, so it could be expected that all of them
are equivalent under transformations of the isotropy subgroup of the
U-duality group. We have shown, however, that in many cases there
are obstructions due to the existence of invariants preserved by the
isotropy subgroup. As a result, the geodesic solutions generically
split into disjoint classes such that the symmetry transformations
act only inside each class, but not between different classes. This
property does not hold for simple cosets like $SL(2,R)/SO(1,1)$ or
$SU(2,1)/S[U(2)\times U(1)]$ corresponding to four-dimensional
Einstein and Einstein-Maxwell theories respectively, but holds for
$SL(2,R)/SO(2,1)$ (five-dimensional vacuum gravity) and for the
coset $G_{2(2)}/((SL(2,R)\times SL(2,R))$ of MSG5 investigated here,
so it presumably is a general feature of large enough cosets. The
deeper group-theoretical significance of the above obstructions also
awaits to be explored.

\section*{Acknowledgments} We wish to thank Paul Sorba for several
enlightening discussions on group theory. AB, CMC and DG would like
to thank LAPTh Annecy-le-Vieux for hospitality at different stages
of this work. DG acknowledges the support of the Russian Foundation
of Fundamental Research under the project 14-02-01092-a. The work of
CMC was supported by the National Science Council of the R.O.C.
under the grant NSC 102-2112-M-008-015-MY3, and in part by the
National Center of Theoretical Sciences (NCTS).

\renewcommand{\theequation}{A.\arabic{equation}}
\setcounter{equation}{0}
\section*{Appendix A: $G_{2(+2)}/((SL(2,R)\times
SL(2,R))$ coset representative}

The $7\times7$ matrix $M$ entering Eq. (\ref{tarmet}) was
constructed in \cite{g2,5to3} and has the symmetrical block
structure:
 \be\lb{Mblock} M =
\left(\begin{array}{ccc} A & B & \sqrt2U \\ B^T & C & \sqrt2V \\
\sqrt2U^T & \sqrt2V^T & S
\end{array}\right)\,,
 \ee
where $A$ and $C$ are symmetrical $3\times3$ matrices, $B$ is a
$3\times3$ matrix, $U$ and $V$ are 3-component column matrices, and
$S$ a scalar. These are given in terms of the moduli by
$$
\begin{array}{l}
A = \left(\begin{array}{cc} \begin{array}{c} \left[(1-y)\lambda +
(2+x)\psi\psi^T  - \tau^{-1}\tom\tom^T\right.\\
\left.+\nu(\psi\psi^T\lambda^{-1}J - J\lambda^{-1}\psi\psi^T)\right]
\end{array} & \tau^{-1}\tom \\
\tau^{-1}\tom^T & -\tau^{-1}
\end{array}\right), \\
B = \left(\begin{array}{cc} (\psi\psi^T-\nu J)\lambda^{-1} -
\tau^{-1}\tom\psi^TJ &
\begin{array}{c}
\left[(-(1+y)\lambda J - (2+x)\nu +
\psi^T\lambda^{-1}\tom)\psi\right.
\\ \left. + (z - \nu J\lambda^{-1})\tom \right]
\end{array} \\
 \tau^{-1}\psi^TJ & -z
\end{array}\right), \\
C = \left(\begin{array}{cc} (1+x)\lambda^{-1} -
\lambda^{-1}\psi\psi^T\lambda^{-1} & \lambda^{-1}\tom-J(z-\nu
J\lambda^{-1})\psi\\ \tom^T\lambda^{-1} +
\psi^T(z+\nu\lambda^{-1}J)J &
\begin{array}{c}
\left[\tom^T\lambda^{-1}\tom - 2\nu\psi^T\lambda^{-1}\tom\right. \\
\left. -\tau(1+x-2y-xy+z^2) \right]
\end{array}
\end{array}\right),
\end{array}
$$\be
\begin{array}{l}
U = \left(\begin{array}{c} (1+x-\nu J\lambda^{-1})\psi -
\nu\tau^{-1}\tom \\ \nu\tau^{-1}
\end{array}\right), \\
V = \left(\begin{array}{c} (\lambda^{-1} + \nu\tau^{-1}J)\psi \\
\psi^T\lambda^{-1}\tom - \nu(1+x-z)
\end{array}\right), \\
S = 1+2(x-y)\,,
\end{array} \lb{coset}
\ee with \be \tom = \omega - \nu\psi\,. \quad x =
\psi^T\lambda^{-1}\psi\,,  \quad y = \tau^{-1}\nu^2\,, \quad z = y -
\tau^{-1}\psi^TJ\tom\,.  \ee

The $7\times7$ matrix representatives $j_M$ of the infinitesimal
generators of $G_{2(+2)}$ may be written in block form
 \be\lb{jgen}
j = \left(\begin{array}{ccc} S & \tilde{V} & \sqrt2U \\
-\tilde{U} & - S^T & \sqrt2V \\ \sqrt2V^T &  \sqrt2U^T & 0
\end{array}\right),
 \ee
where $S$ is a $3\times3$ matrix, $U$ and $V$ are 3-component column
matrices, $U^T$ and $V^T$ the corresponding transposed row matrices,
and $\tilde{U}$, $\tilde{V}$ are the $3\times3$ dual matrices
$\tilde{U}_{ij} = \epsilon_{ijk}U_k$. The matrices ${m_a}^b$, $n^a$
and $\ell_a$ generating the vacuum $SL(3,R)$ subgroup of $G_{2(+2)}$
are of type $S$, the corresponding $3\times3$ blocks being
 \ba
S_{{m_0}^0}\!\! &=&\!\! \left(\begin{array}{ccc}1&0&0\\0&0&0\\0&0&-1
\end{array}\right),\;
S_{{m_0}^1} =
\left(\begin{array}{ccc}0&1&0\\0&0&0\\0&0&0\end{array}\right) ,\;\nn\\
S_{{m_1}^0} &=&
\left(\begin{array}{ccc}0&0&0\\1&0&0\\0&0&0\end{array}\right) ,\;
S_{{m_1}^1} =
\left(\begin{array}{ccc}0&0&0\\0&1&0\\0&0&-1\end{array}\right)
,\;\nn
\\ S_{n^0}\! \!&=&\!\!
\left(\begin{array}{ccc}0&0&0\\0&0&0\\-1&0&0
\end{array}\right),\;
S_{n^1} =
\left(\begin{array}{ccc}0&0&0\\0&0&0\\0&-1&0\end{array}\right),\;\\
S_{\ell_0} &=&
\left(\begin{array}{ccc}0&0&1\\0&0&0\\0&0&0\end{array}\right),\;
S_{\ell_1} =
\left(\begin{array}{ccc}0&0&0\\0&0&1\\0&0&0\end{array}\right).\nn
 \ea
The matrices $p_a$ and $q$ are of type $U$, the corresponding
$1\times3$ blocks being \be U_{p_0} =  \left(\begin{array}{c} 1 \\ 0
\\ 0
\end{array}\right),\;\;
U_{p_1} =  \left(\begin{array}{c} 0 \\ 1 \\ 0
\end{array}\right),\;\;
U_{q} =   \left(\begin{array}{c} 0 \\ 0 \\ -1
\end{array}\right).
\ee The matrices $r^a$ and $t$ are of type $V$,  the corresponding
$1\times3$ blocks being \be V_{r^0} =  \left(\begin{array}{c} 1 \\ 0
\\ 0
\end{array}\right),\;\;
V_{r^1} =  \left(\begin{array}{c} 0 \\ 1 \\ 0
\end{array}\right),\;\;
V_{t} =   \left(\begin{array}{c} 0 \\ 0 \\ 1
\end{array}\right).
\ee

\renewcommand{\theequation}{B.\arabic{equation}}
\setcounter{equation}{0}
\section*{Appendix B: Proof that the 3-G\"odel solution cannot be
transformed to the Schwarzschild black string}

The fact that the 3-G\"odel solution (\ref{bgods}) and the
Schwarzschild black string (\ref{blackS}) have the same
three-dimensional reduced metric (\ref{3red}) suggests that their
matrix representatives might be related by a $G_{2(+2)}$
transformation,
 \be\lb{stog}
M_G = P_{SG}^T M_S P_{SG}\,,
 \ee
the corresponding constant matrices $\eta$ and $\A$ being related by
 \be\lb{transfea}
\eta_G = P_{SG}^T \eta_S P_{SG}\,, \quad \A_G =
P_{SG}^{-1}\A_SP_{SG}\,.
 \ee
We prove here that this is impossible.

We first consider the second equation (\ref{transfea}). The
Schwarzschild matrix $\A_S = \mbox{\rm diag} (1,0,-1,-1,0,1,0)$ has
the three degenerate eigenvalues $\pm1$ and $0$ with the obvious
eigenvectors $({\psi}_{Si_{\pm}})^a = \delta_{i_{\pm}}^a$ and
$(\psi_{Si_0})^a = \delta_{i_0}^a$. The matrix $\A_G$ has the same
degenerate eigenvalues $\pm1$ and $0$ with suitably orthonormalized
eigenvectors $(\psi_{Gi_{\pm}})^a$ and $(\psi_{Gi_0})^a$. The
similarity transformation, given by the sum $P_{SG} =
\psi_{Sk_\alpha}\psi_{Gk_\alpha}^T$, is thus
 \be
{(P_{SG})^a}_b = (\psi_{Ga})^b\,.
 \ee
Now let us compute, from the first equation (\ref{transfea}),
 \be
(\eta_G)_{77} = (\eta_S)_{ab}(\psi_{Ga})^7(\psi_{Gb})^7 =
[(\psi_{Gi_0})^7]^2 - [(\psi_{Gi_+})^7]^2 - [(\psi_{Gi_-})^7]^2
 \ee
(with sum over repeated indices implied). Remembering that
$\psi_{Gi_0}$ solves $\A_G\psi_{Gi_0} = 0$, we find from the second
and fourth line of (\ref{ag}) that $(\psi_{Gi_0})^7 = 0$, leading to
$(\eta_G)_{77} < 0$, in contradiction with (\ref{etag}).

\end{document}